\documentclass[12pt]{iopart}
\usepackage[dvips]{graphicx}
\begin{document}
\title[Conductivity of a superlattice]
{Conductivity of a superlattice with parabolic miniband}
\author{G M Shmelev$^1$, I~I~Maglevanny$^1$
and E~M~Epshtein$^2$
      }
\address{$^1$
Volgograd State Pedagogical University, 400131, Volgograd, Russia
        }
\address{$^2$
Institute of Radio Engineering and Electronics, Fryazino, 141190, Russia
        }
\ead{shmelev@fizmat.vspu.ru}

\begin{abstract}
The static and high-frequency differential conductivity
of a one-dimensional superlattice  with parabolic miniband, in which
the dispersion law is
assumed to be parabolic up to the Brillouin zone edge,
are investigated theoretically. Unlike the earlier published
works, devoted to this problem, the
novel formula for the static current density
contains temperature dependence, which leads to
the current maximum shift to the low field side with increasing temperature.

The high-frequency differential conductivity response
properties including the temperature dependence is examined
and
opportunities of creating
a terahertz oscillator on Bloch electron oscillations in such superlattices
are discussed.

Analysis shows that superlattices with
parabolic miniband dispersion law may be used for
generation and amplification of terahertz fields
only at very low temperatures ($T\to0$).
\end{abstract}

%\submitto{\JPA}
\maketitle

\section{Introduction}\label{sec1}
In present work, we study theoretically
the static and high-frequency conductivity
of a semiconductor superlattice (SL). Unlike the earlier published
numerous works, devoted to this problem, where the conventional cosine-type
model was used for the conduction miniband, here a dispersion law is
considered in form of a truncated parabola, i.~e. the dispersion law is
assumed to be parabolic up to the Brillouin zone edge.
Such a problem statement is of interest, among others,
from the view point of opportunities of creating
a terahertz oscillator on Bloch electron oscillations in SLs.
In works~\cite{Romanov1}-\cite{Romanov5} different variants
of realization of such opportunity were discussed and it was mentioned that
the main obstacle consists in using
the non-optimal SL structures, in particular, the SL with
cosine-type miniband.

Thus the theoretical investigations of electric properties of SLs with
other dispersion laws are necessary,
all the more so since the modern technology allows to vary
widely the form of the potential relief and the SL energy spectrum.

The main condition for realization of Bloch oscillator consists in
existence
of negative high-frequency differential conductivity on that regions of
current-voltage characteristic where the static
differential conductivity is positive. In~\cite{Romanov2}
it was shown that this condition holds, in particular,
in SL with parabolic miniband. But this result was obtained
in the limiting case $T\to0$. Here we find the
temperature dependence of conductivity of such SL and define the
temperature criterion by which the mentioned condition
holds practically.

This article is structured as follows.
In Section~\ref{sec2} we derive an expression
for static conductivity of SL with
parabolic miniband, which is valid for any temperatures.
In Section~\ref{sec3} we derive the corresponding expression
for high-frequency differential conductivity.
Section~\ref{sec4} presents the conclusions of our work.

\section{Static distribution function and current-voltage characteristic}
\label{sec2}

The electron energy in the SL lowest miniband is~\cite{Romanov1}
\begin{equation}\label{1}
  \varepsilon(\mathbf p)= \varepsilon(\mathbf p_\bot)
  +\frac{p^2}{2m},\quad -\frac{\pi\hbar}{d}<p<\frac{\pi\hbar}{d},
\end{equation}
where $\mathbf p$ is quasimomentum, $d$ is SL period, $x$ axis being
directed along the SL axis, $\varepsilon(\mathbf p_\bot)$ is in-plane
electron energy, $\pi^2\hbar^2/md^2\equiv\Delta$ is double miniband width,
$m$ is effective electron mass.

In quasi-classical situation ($\Delta\gg eEd,\,\hbar/\tau$, where $\tau$ is
electron momentum relaxation time, $e$ is electron charge), the current
density in electric field
$\mathbf E^{tot}(t)$ may be found by solving
Boltzmann equation with collision integral within $\tau$-approximation:
\begin{equation}\label{2_}
\frac{\partial F(\mathbf p,t)}{\partial t} +
\left(e\mathbf E^{tot}(t),\frac{\partial F(\mathbf p,t)}
{\partial\mathbf p}\right)
  =\frac{F_0(\mathbf p)- F(\mathbf p,t)}{\tau},
\end{equation}
where $F_0(\mathbf p)$ is equilibrium electron distribution function,
$F(\mathbf p,t)$ is unknown distribution function perturbed due the
electric field.
Below we  use dimensionless variables
by changing $\mathbf pd/(\pi\hbar)\rightarrow\mathbf p$,
$\mathbf E^{tot}ed\tau/(\pi\hbar)\rightarrow\mathbf E^{tot}$,
$T/\Delta\rightarrow T$, $t/\tau\rightarrow t$
($T$ is temperature in energy units).

With the field  $\mathbf E^{tot}(t)$ is directed along the SL axis
$\left(\mathbf E^{tot}(t)=\left(E^{tot}(t),0,0\right)\right)$, we have
$F(\mathbf p,t)=f_0(\mathbf p_\bot)f(p,t)$, $F_0(\mathbf p)=f_0(\mathbf
p_\bot)f_0(p)$, where $f_0(p)$ is equilibrium distribution function,
normalized to the carrier density
$n$ ($f_0(\mathbf p_\bot)$ being normalized to unity).
Thus, the function  $f(p,t)$ satisfies the following equation
\begin{equation}\label{3_}
\frac{\partial f(p,t)}{\partial t} +
  E^{tot}(t)\frac{\partial f(p,t)}{\partial p}=f_0(p)-f(p,t),\quad (-1<p<1).
\end{equation}
with periodicity conditions $f(1,t)=f(-1,t)$.

In a static field $E^{tot}(t)=E=const$, and
denoting $f(p)=f_c(p,E,T)$, we get
\begin{equation}\label{3}
  E\frac{{\rm d} f_c}{{\rm d} p}=f_0-f_c,\quad (-1<p<1).
\end{equation}

We consider non-degenerate electron gas, so that
\begin{equation}\label{4}
  f_0(p,T)=2n\left[\sqrt{2\pi T}\mathrm{erf}
  \left(\frac{1}{\sqrt{2T}}\right)\right]^{-1}\exp
\left(-\frac{p^2}{2T}\right)
\end{equation}
where $\mathrm{erf}(z)$ is error function.
In the low temperature limit ($T\rightarrow 0$)
the relation~(\ref{4}) reduces to
the function used in~\cite{Romanov1}: $g_0(p)=2n\delta(p)$.

The exact solution of~(\ref{3}) with periodicity condition,
$f_c(-1)=f_c(1)$, takes the form~\cite{Shmelev3}
\begin{eqnarray}
\fl
f_c(p,E,T)=\frac{n}{E\mathrm{erf}\left(1/\sqrt{2T}\right)}
\exp\left(\frac{T}{2E^2}-\frac{p}{E}\right)
  \left\{\mathrm{erf}\left(\frac{p}{\sqrt{2T}}-\frac{\sqrt
  T}{\sqrt{2}E}\right)\right.\nonumber\\
-\left[\exp\left(\frac{2}{E}\right)-1\right]^{-1}
\mathrm{erf}\left(\frac{\sqrt T}
{\sqrt{2}E}-\frac{1}{\sqrt{2T}}\right)\nonumber \\
+\left.\left[1-\exp\left(-\frac{2}{E}\right)\right]^{-1}
\mathrm{erf}\left(\frac{\sqrt
  T}{\sqrt{2}E}+\frac{1}{\sqrt{2T}}\right)\right\},\quad
  -1<p<1.
\label{7}
\end{eqnarray}

In limiting case $E\rightarrow 0$~(\ref{7}) reduces to~(\ref{4}).
In another limiting case, $T\rightarrow 0$,
we get the distribution function found in~\cite{Romanov1}:
\begin{equation}\label{8}
\fl
  g(p,E)=\frac{2n}{E}\exp\left(-\frac{p}{E}\right)
\cases{[1-\exp(-2/E)]^{-1},
   & $0<p<1$,\cr
[\exp(2/E)-1]^{-1}, & $-1<p<0$}.
\end{equation}

\begin{figure}[ht]
\begin{center}
\includegraphics[width=85mm,height=80mm]{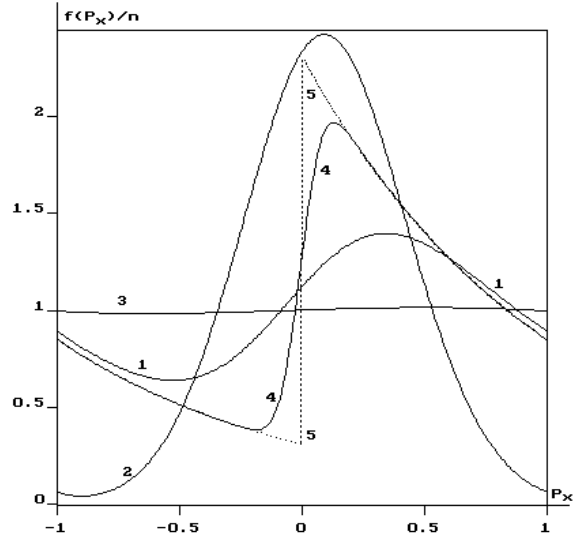}
\end{center}
\caption{Distribution function $f_c(p)$ at various values
of the driving field and
temperature. 1) $E=1,\;T=0.1$; 2)  $E=0.1,\;T=0.1$;
3) $E=2,\;T=2$; 4) $E=1,\;T=0.005$. The dashed
curve 5 represents function $g(p)$ at $E=1$.}
\label{fig1}
\end{figure}

The function~(\ref{7}) satisfies the same normalization condition as the
equilibrium function $f_0$
\begin{equation}\label{9}
  \frac{1}{2}\int\limits_{-1}^1f_c(p,E,T)\,dp=n
\end{equation}
and, therefore, it makes the integral of right-hand side of formula~(\ref{3})
vanish. Besides, the integral of
left-hand side of the Boltzmann equation~(\ref{3}) vanishes too, because
of the periodicity condition mentioned.  The
distribution function $f_c(p,E,T)$ at several values of $E$ and $T$ is shown
in figure~\ref{fig1}.

The current density $j$ in the direction of SL axis
can be found (in dimensional units) by a conventional way
\begin{equation}\label{10}
  j=\frac{ed}{2\pi\hbar m}\int\limits_{-\pi\hbar /d}^
{\pi\hbar /d}pf_c(p)\,dp.
\end{equation}

By substitution function~(\ref{7}) into~(\ref{10}) we get
\begin{eqnarray}
\fl
j(E,T)=E+
\left[2\mathrm{erf}\left(\frac{1}{\sqrt{2T}}\right)
\sinh\left(\frac{1}{E}\right)
\right]^{-1}  \exp\left(\frac{T}{2E^2}\right)\nonumber\\
\times
\left[\mathrm{erf}\left(\frac{\sqrt{T}}{E\sqrt{2}}-
\frac{1}{\sqrt{2T}}\right) -
\mathrm{erf}\left(\frac{\sqrt{T}}{E\sqrt{2}}+
\frac{1}{\sqrt{2T}}\right)
\right].
\label{10_}
\end{eqnarray}
Here
$j$ is expressed in units of $j_0=ne\Delta d/\pi\hbar$, while all
the quantities are written in dimensionless form.

Equation~(\ref{10_})
determines the current-voltage characteristic for the parabolic miniband SL
with the current density temperature dependence taking into account.

To warrant numerical stability we present formula~(\ref{10_})
in the following form
\begin{equation}\label{11}
\fl
j(E,T)=E\sigma(E,T),\quad \sigma(E,T)=1-\sqrt{\frac{2}{\pi T}}
\frac{\exp(-0.5/T)+A(E,T)}{\mathrm{erf}(1/\sqrt{2T})},
\end{equation}
where $\sigma(E,T)$ is the conductivity and
\begin{equation}\label{11_}
A(E,T)=\frac{E^2}{T\sinh (1/E)}
\int\limits_0^{1/E}\exp\left(-\frac{s^2E^2}{2T}\right) s\sinh s\,ds.
\end{equation}
The value of $A(E,T)$ can be estimated numerically with high accuracy.

Expanding the exponent in a power series we get
\begin{equation}\label{11___}
A(E,T)=\frac{1}{T}\sum_{n=0}^\infty\frac{(-1)^n}{(2n)!!}
\frac{G_n(E)}{T^n},
\end{equation}
where functions $G_n(E)$ are defined by recurrent formula
\begin{equation}\label{11____}
\fl
G_0=E\coth\left(\frac{1}{E}\right)-E^2,\quad
G_n=G_0+2nE^2\left[(2n+1)G_{n-1}-1\right].
\end{equation}

As $G_n(E)\in[0,1/(2n+3))$, series~(\ref{11___})
converges quickly. As numerical experiments show, first four
terms of series~(\ref{11___}) give good approximation at $T>0.5$.

At $|E|\to0$ we have $A(E,T)\to0$, so
in low fields ($|E|\ll 1$) in linear approximation on $E$ we have
\begin{equation}\label{13}
j(E,T)=E\left(1-\sqrt{\frac{2}{\pi T}}
\frac{\exp(-0.5/T)}{\mathrm{erf}(1/\sqrt{2T})}\right)
=E\left\langle\frac{p^2}{T}\right\rangle_0,
\end{equation}
where angle brackets mean averaging over the equilibrium distribution.
Note that the conductivity temperature dependence in low fields (the
expression within round brackets in~(\ref{13})) is close to the
analogous dependence for the miniband cosine model
($I_1(1/2T)/I_0(1/2T)$, $I_n(z)$ being the modified Bessel function).

In high fields ($|E|>1$) we have
\begin{equation}\label{_13}
\fl
\sigma(E,T)\approx \frac{\sqrt{2/\pi}}{E^2T\sqrt{T}
\mathrm{erf}(1/\sqrt{2T})}
\sum_{n=0}^\infty\frac{(-1)^{n+1}}{(2n)!!}\frac{D_n}{T^n},
\quad D_n=\sum_{k=2}^{n+2}\frac{2^{2k}(2n+1)!B_{2k}}
{(2k)!(2n-2k+5)!},
\end{equation}
where $B_m$ are Bernoulli numbers.

For low temperatures ($T\ll1$), using~(\ref{11_}), we get
\begin{equation}
\fl
\sigma(E,T)\approx 1-\frac{1}{\mathrm{erf}\left(1/\sqrt{2T}\right)}
\left[\frac{1}{E\sinh(1/E)}\exp\left(\frac{T}{2E^2}\right)
+\sqrt{\frac{2}{\pi T}}\exp\left(-\frac{1}{2T}\right)\right].
\label{_12_}
\end{equation}
As numerical experiments show,
formula~(\ref{_12_}) gives good approximation at $T<0.07$.
In limiting case $T\to0$ from~(\ref{_12_}) we get the expression
that was found in~\cite{Romanov1}
\begin{equation}\label{12}
j=j(E)=E-\frac{1}{\sinh(1/E)}.
\label{__12}
\end{equation}

\begin{figure}[ht]
\begin{center}
\includegraphics[width=85mm,height=80mm]{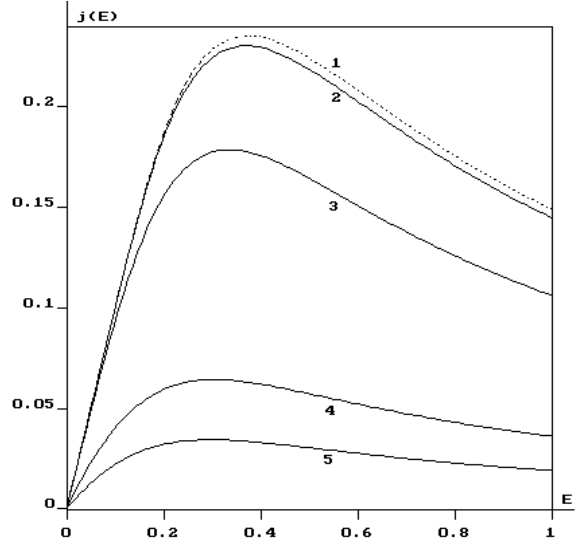}
\end{center}
\caption{Current-voltage characteristic at different values of temperature.
1) $T=0$; 2)  $T=0.01$;
3) $T=0.1$; 4) $T=0.5$; 5) $T=1$.
}
\label{fig2}
\end{figure}

From~(\ref{13},\ref{_13}) it follows that $j\sim E$ at $|E|\ll1$
and $j\sim 1/E$ at $|E|\gg1$. Therefore at fixed
temperature $T=fix$ the function
$j(E,T)$ reaches its maximum at some value $E=E_{C}(T)>0$
and negative differential conductivity is realized at $E>E_{C}(T)$
(see figure~\ref{fig2}).

Note, that $E_{C}(T)$ decreases with increasing temperature.
Essentially, that $E_{C}$ value does not depend on the temperature at all
in the cosine model: $E_{C}=1/\pi\approx0.318$.

The parametric representation of dependence
$E_{C}(T)$ is defined by equation $\sigma_{ d}=0$,
where $\sigma_{ d}=\partial j/\partial E$
is the differential conductivity. Using~(\ref{11},\ref{11_}), we get
\begin{eqnarray}
\sigma_{ d}(E,T)=&1+\frac{1}{E^2}\left\{
\left[E\coth\left(\frac{1}{E}\right)-T\right]
\left[\sigma(E,T)-1\right]\right.-\nonumber\\
&\left.\sqrt{\frac{2T}{\pi}}
\frac{1}{\mathrm{erf}\left(1/\sqrt{2T}\right)}
\exp\left(-\frac{1}{2T}\right)\right\},
\label{_f2}
\end{eqnarray}

Thus function $E_{C}(T)$ is defined implicitly by equation
\begin{equation}\label{f2_}
\fl
E^2\sqrt{\frac{\pi T}{2}}\mathrm{erf}\left(\frac{1}{\sqrt{2T}}\right)
+TA(E,T)=E\coth\left(\frac{1}{E}\right)\left[A(E,T)
+\exp\left(-\frac{1}{2T}\right)\right],
\end{equation}
and it is sufficient to solve this equation at $E>0$.

\begin{figure}[ht]
\begin{center}
\includegraphics[width=85mm,height=80mm]{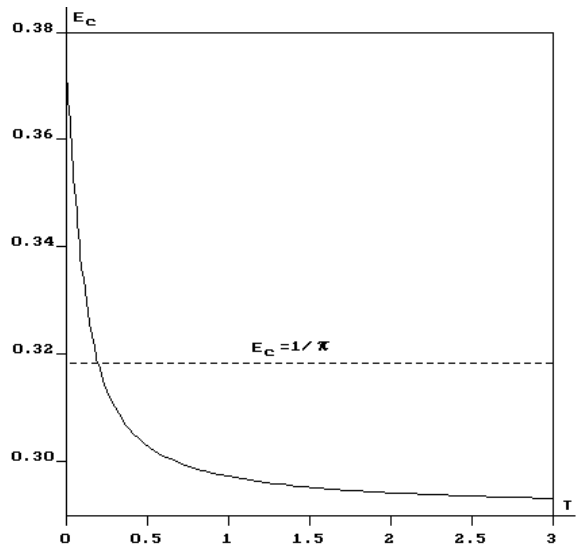}
\end{center}
\caption{The dependence $E=E_C(T)$.
The dashed curve $E_C=1/\pi$ represents
$E_C$ for cosine model.
}
\label{fig3}
\end{figure}

The numerical solution of equation~(\ref{f2_}) at $E$ versus $T$
is presented in figure~\ref{fig3}.

Note that dependence $E=E_{C}(T)$ is monotone so the inverse
function $T_C=T_C(E)$ exists. To investigate behavior of function
$T_C(E)$, consider first the case of high temperatures $T\gg1$.
Expanding all functions in a power series on $1/T$ and neglecting
all terms $o(1/T^2)$, we get
\begin{equation}\label{appr2}
\fl
T_C\approx \frac{(45E^4+22.5E^2+1.8)\tanh^2\left(1/E\right)-
(36E+3E)\tanh\left(1/E\right)-9E^2-1.5}
{(9E^2+4)\tanh^2\left(1/E\right)-6E\tanh\left(1/E\right)-3}
\end{equation}
By that
\begin{equation}\label{appr3}
\lim_{E\to E_1+0}T_C(E)=+\infty,
\end{equation}
where $E_1\approx0.29104955$ is the root of equation
\begin{equation}\label{appr3_}
(9E^2+4)\tanh^2\left(\frac{1}{E}\right)-
6E\tanh\left(\frac{1}{E}\right)-3=0.
\end{equation}

Consider now the case of low temperatures $T\ll1$.
Using~(\ref{_12_}), we get
\begin{equation}\label{appr5}
\fl
T_C(E)\approx 2E^2\left[E^2\tanh\left(\frac{1}{E}\right)
\sinh\left(\frac{1}{E}\right)-1\right]
\left[1-2E\tanh\left(\frac{1}{E}\right)\right]^{-1}.
\end{equation}
By that
\begin{equation}\label{appr6}
\lim_{E\to E_2-0}T_C(E)=0,
\end{equation}
where $E_2\approx0.373681745$ is the root of equation
\begin{equation}\label{appr7}
E^2\tanh\left(\frac{1}{E}\right)
\sinh\left(\frac{1}{E}\right)=1.
\end{equation}

Therefore function $E_{C}(T)$ is defined for $T>0$ and
\begin{equation}\label{appr8}
\lim_{T\to 0}E_{C}(T)=E_2,\quad \lim_{T\to +\infty}E_{C}(T)=E_1.
\end{equation}

\section{High-frequency differential conductivity}\label{sec3}
In this section we will determine the induced superlattice current
in the presence of an external electric field given by
\begin{equation}
E^{tot}(t)=E+E_0\cos\omega t,
\label{field}
\end{equation}
where $\omega$ is measured in units of $\tau^{-1}$.
Within the scope of quasi-classical conditions the value of
$E$ is arbitrary. Assuming the amplitude of variable field
$E_0$ to be much smaller then the static field $E$,
consider the time-dependent field in linear approximation.
The distribution function may be found in a form
\begin{equation}\label{add29}
f(p,E,T,t)=f_c(p,E,T)+f_1(p,E,T,\omega)\exp(-i\omega t),
\end{equation}
here $f_1(p,E,\omega)$ satisfies the following equation~\cite{Romanov2}
\begin{equation}
E\frac{\partial f_1}{\partial p}+(1-i\omega)f_1=
-E_0\frac{\partial f_c}{\partial p}
\label{add30}
\end{equation}
with periodicity condition $f_1(-1,E,\omega)=f_1(1,E,\omega)$
and by
\begin{equation}
\int_{-1}^{1}f_1(p,E,T,\omega)\,dp=0.
\label{add31}
\end{equation}

It is easily to show that required solution is
\begin{equation}
f_1(p,E,T,\omega)=\frac{i}{\omega}\cdot\frac{E_0}{E}
\left[f_c(p,E,T)+f_c\left(p,\frac{E}{1-i\omega}\right),T\right].
\label{f1}
\end{equation}
With the help of~(\ref{f1}) the dynamic (high-frequency)
differential conductivity  can be found by a conventional way. The result is
\begin{equation}
\sigma_1(E,T,\omega)=\frac{i}{\omega E}\left[
j(E,T)-j\left(\frac{E}{1-i\omega},T\right)
\right].
\label{romanov1}
\end{equation}

From~(\ref{romanov1}) it follows that at
$\omega\to0$ the value $\sigma_1(E,T,\omega)$ tends to
static differential conductivity~(\ref{_f2})
\begin{equation}
\lim_{\omega\to0}\sigma_1(E,T,\omega)=\sigma_{d}(E,T).
\label{romanov2}
\end{equation}

Using~(\ref{10_}), we get
\begin{equation}
{{Re}}\,\sigma_1(E,T,\omega)=\frac{i}{2\omega E}\left[
j\left(\frac{E}{1+i\omega},T\right)-j\left(\frac{E}{1-i\omega},T\right)\right].
\label{Realsigma1}
\end{equation}
For numerical computations we present expression~(\ref{Realsigma1}) in a form
\begin{eqnarray}
{{Re}}\,\sigma_1(E,T,\omega)=\frac{1}{1+\omega^2}
\nonumber \\
\fl
-\sqrt{\frac{2}{\pi T}}\frac{\left[\sinh^2(1/E)+\sin^2(\omega/E)\right]^{-1}}
{\omega E\mathrm{erf}\left(1/\sqrt{2T}\right)}
\left[\cosh\frac{1}{E}\sin\frac{\omega}{E}
\int_0^1\exp\left(-\frac{s^2}{2T}\right)
\cosh\frac{s}{E}\cos\frac{s\omega}{E}\,ds\right.
\nonumber \\
-\left.\sinh\frac{1}{E}\cos\frac{\omega}{E}
\int_0^1\exp\left(-\frac{s^2}{2T}\right)
\sinh\frac{s}{E}\sin\frac{s\omega}{E}\,ds
\right].
\label{Realsigma}
\end{eqnarray}

At $T\to0$ from~(\ref{Realsigma}) we get the expression presented
in~\cite{Romanov2}
\begin{equation}
{{Re}}\,\sigma_1(E,0,\omega)
=\frac{1}{1+\omega^2}-\frac{\cosh(1/E)\sin(\omega/E)}
{\omega E\left[\sinh^2(1/E)+\sin^2(\omega/E)\right]}
\label{Realsigma0}
\end{equation}

The opportunities of creating a terahertz oscillator on Bloch
electron oscillations in SLs are defined by conditions of existence
of negative high-frequency differential conductivity on that regions of
current-voltage characteristic where the static
differential conductivity is positive~\cite{Romanov2}-\cite{Romanov5}.
These conditions would prevent development of undesirable
domain instabilities (Gunn effect).

Let $\Omega=eEd/\hbar$ be the Bloch oscillations frequency
which in normalized measurement units is equal to $\pi E$.
Then the static differential conductivity $\sigma_{d}$ is positive at
$\Omega<\Omega_C$ and negative at  $\Omega>\Omega_C$,
where $\Omega_C=\pi E_C(T)\in(0.914,1.174)$. Thus the conditions
of low-frequency domain instability suppression are defined by that
values of parameters $\omega$ and $\Omega$, for which
\begin{equation}
\cases{
\Omega<\Omega_C\cr
\sigma_1(\Omega,T,\omega)<0
       }
\label{domain}
\end{equation}

The existence of such conditions
for regarded model of dispersion law
was discovered in~\cite{Romanov2} in limiting case $T\to0$.

\begin{figure}[!htb]
\begin{center}
\unitlength=1mm
\begin{picture}(160,75)
\put(39,-5){a)} \put(120,-5){b)}
\includegraphics[width=75mm,height=75mm]{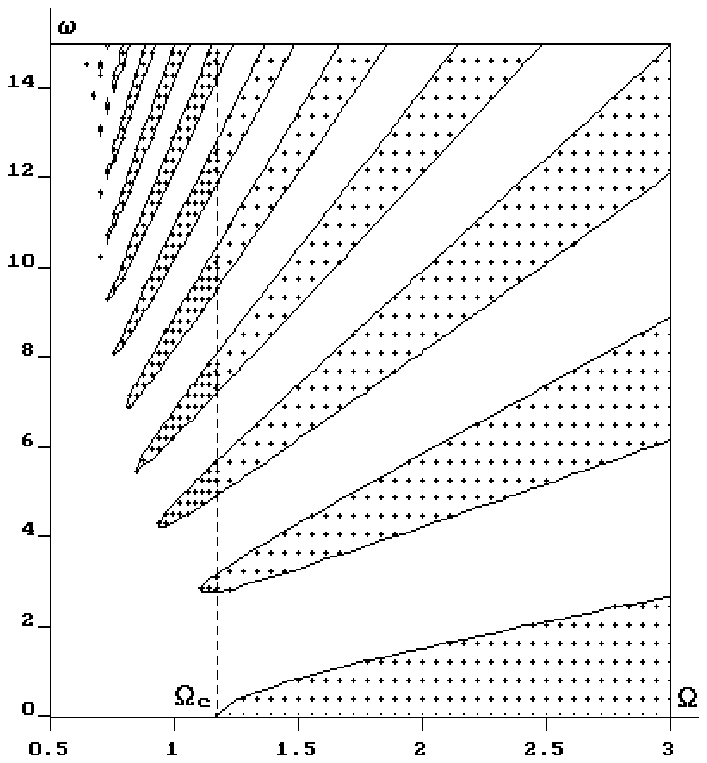}
\quad
\includegraphics[width=75mm,height=75mm]{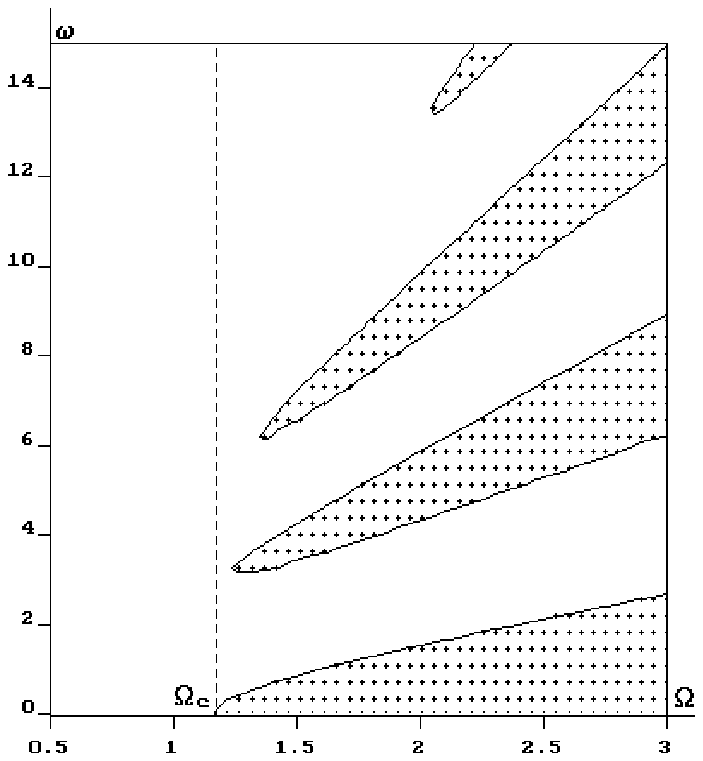}
\end{picture}
\end{center}
\caption{The regions of negative high-frequency differential conductivity
at parameter plane $(\Omega,\omega)$.
a) $T\to0$, $\Omega_C=1.174$.
b) $T=0.01$, $\Omega_C=1.06$.
        }
\label{fig56}
\end{figure}

But conditions~(\ref{domain}) prove to be very sensitive to temperature
increasing.
In figure~\ref{fig56} the regions in parameter space
$(\Omega,T,\omega)$ are presented in which the
high-frequency differential conductivity is negative.

The boundary lines of these regions are defined by condition
${{Re}}\,\sigma_1(E,T,\omega)=0$. At these lines we have
$\omega\approx k\Omega$, $k=1,2,\dots$. Thus
the frequencies at which the
high-frequency differential conductivity changes sign
are multiples of the Bloch frequency.

Note the existence of regions of
low-frequency domain instability suppression at $T=0$ and
absence of such regions at $T=0.01$.

\begin{figure}[!htb]
\begin{center}
\unitlength=1mm
\begin{picture}(160,75)
\put(5,-5){a)} \put(70,-5){b)}
\includegraphics[width=75mm,height=75mm]{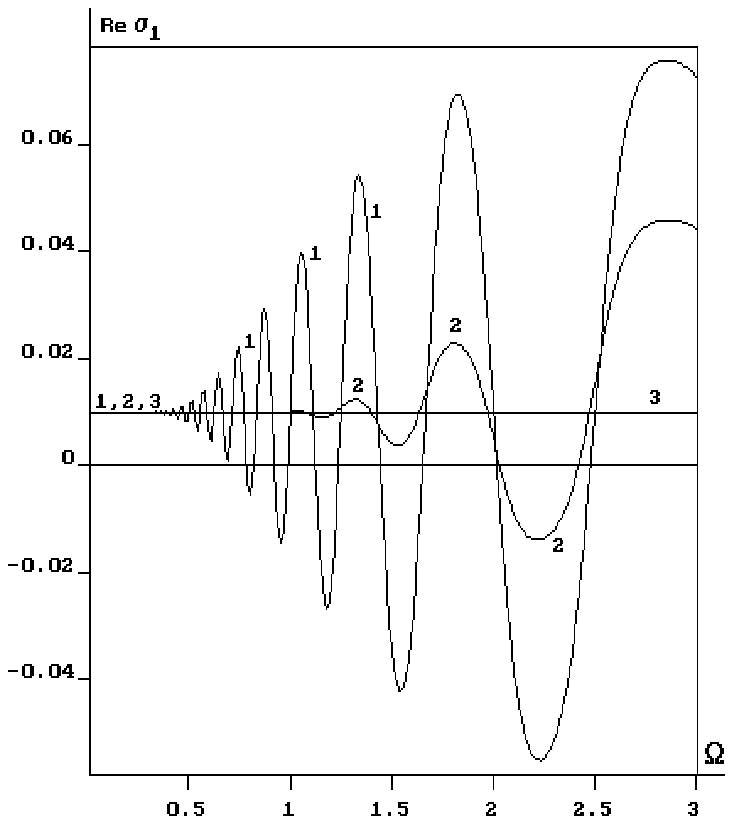}
\quad
\includegraphics[width=75mm,height=75mm]{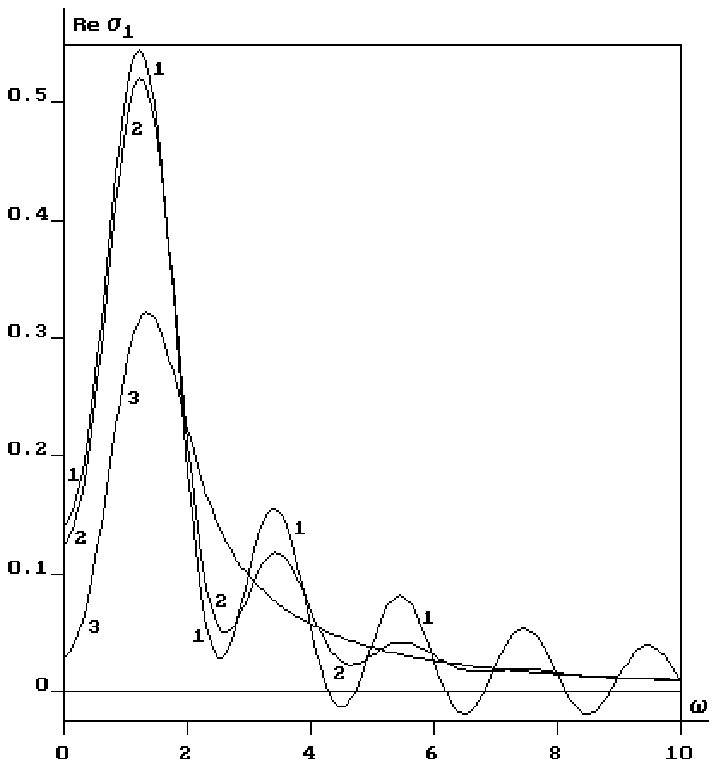}
\end{picture}
\end{center}
\caption{
a) Driving field dependence of high-frequency
differential conductivity at $\omega=10$.
1) $T=0$, $\Omega_C=1.174$;
2) $T=0.01$, $\Omega_C=1.163$; 3) $T=0.1$, $\Omega_C=1.06$.
b) Dependence of high-frequency
differential conductivity on $\omega$ at $\Omega=1$.
1) $T=0$; 2) $T=0.01$; 3) $T=0.1$.
At such temperatures the static differential conductivity
$\sigma_d=\left.\sigma_1\right|_{\omega=0}$ is positive.
        }
\label{fig47}
\end{figure}

The dependence of function $\sigma_1$ on parameters $\Omega$
and $\omega$ is presented in figure~\ref{fig47}.

Note that by temperature increasing the oscillations of $\sigma_1$
become suppressed at $\Omega<\Omega_c$
and negative high-frequency differential conductivity disappears.

\section{Conclusion}\label{sec4}
In present paper, an exact distribution function has been found of the
carriers in the lowest parabolic miniband of a SL,
placed in the dc electric field, parallel to SL axis.
The novel formula for the static current density in SL
contains temperature dependence, which leads to
the current maximum shift to the low field side with increasing temperature.

We have obtained explicit expression for high-frequency
differential conductivity at arbitrary temperature. It was shown
that high-frequency differential conductivity is very sensitive to
temperature of SL. We have compared high-frequency electron
behavior at different temperatures and exhibited the drastic
change in the character of regions where the high-frequency
differential conductivity is negative. In particular we have
discovered that the possibility of low-frequency domain
instability suppression may be realized only at $T\to0$.

In summary, our analysis shows that SLs with
parabolic miniband dispersion law may be used for
generation and amplification of terahertz fields
only at very low temperatures ($T<0.01\Delta$).

The numerical estimations
of the effects predicted are reduced, in general, to
measurement units of electric field and temperature. At $d=10^{-7}$ cm,
$\tau=10^{-12}$ s, $\Delta\approx10^{-2}$ eV we get that units for $E$ and
$T$ are $\approx2\cdot10^3$ and $\approx100$ K respectively.
Thus the condition $T<0.01\Delta$ is equivalent to $T<1$ K.


\section*{References}
\begin{thebibliography}{10}
\bibitem{Romanov1}
Romanov Yu A  2003 {\it Phys. Solid State} {\bf 45} 559
\bibitem{Romanov2}
Romanov Yu A, Mourokh L G and Horing N J M 2002 {\it cond-mat/0209365}
\bibitem{Romanov4}
Romanov Yu A and Romanova J Yu 2004 {\it Phys. Solid State} {\bf 46} 164
\bibitem{Romanov5}
Romanov Yu A and Romanova J Yu 2005 {\it Phys. Semicond. } {\bf 39} 147
\bibitem{Shmelev3}
Shmelev~G~M, Epshtein~E~M and Gorshenina T A 2005 {\it cond-mat/0503092}
\end{thebibliography}
\end{document}